\documentclass[11pt]{article}
\usepackage{amsmath,amssymb}

\def\qed{$\Box$}

\newcommand\comment[1]{}

\newtheorem{thm}{Theorem}[section]
\newtheorem{prop}[thm]{Proposition}

\newtheorem{cor}[thm]{Corollary}

\newcommand{\qedsymb}{\hfill{\rule{2mm}{2mm}}}

\def\ra{\rangle}

\newcommand{\be}{\begin{eqnarray}}
\newcommand{\ee}{\end{eqnarray}}

\renewcommand{\epsilon}{\varepsilon}

\begin{document}

\title{\Large \bf The hidden subgroup problem and permutation group theory}

\author{
Julia Kempe\\
UC Berkeley, Berkeley, CA94720 and \\
CNRS \& LRI, Universit\'e Paris-Sud\\
91405 Orsay Cedex, France\\
kempe@lri.fr
 \and
Aner Shalev\\
Institute of Mathematics\\
The Hebrew University\\
Jerusalem 91904, Israel\\
shalev@math.huji.ac.il
}

\date{\today}

\maketitle

\begin{abstract}
We employ concepts and tools from the theory of finite permutation
groups in order to analyse the Hidden Subgroup Problem via Quantum
Fourier Sampling (QFS) for the symmetric group. We show that under
very general conditions both the weak and the random-strong form
(strong form with random choices of basis) of QFS fail to provide any
advantage over classical exhaustive search. In particular we give a
complete characterisation of polynomial size subgroups, and of
primitive subgroups, that can be distinguished from the identity
subgroup with the above methods.  Furthermore, assuming a plausible
group theoretic conjecture for which we give supporting
evidence, we show that weak and random-strong QFS for the symmetric
group have no advantage whatsoever over classical search.
\end{abstract}

\section{Introduction}

In the last decade quantum computation has provided us with powerful tools 
to solve problems not known to be classically efficiently solvable, like  
factoring \cite{Shor:94a} and discrete log \cite{Kitaev:95a}. Nearly all 
the problems in which a quantum computer excels more than quadratically with respect to its 
classical counterpart can be cast into the framework of the Hidden Subgroup 
Problem (HSP).  Let $G$ be a finite group and $H \le G$ a subgroup. 
Given a function 
$f: G \rightarrow S$ that is constant on (left)-cosets $gH$ of $H$ and 
takes different values for different cosets, determine a set of generators 
for $H$. The decision version of this problem is to determine whether there 
is a non-identity hidden subgroup or not. 

The reason that quantum computers seem to provide a speed-up for this
type of problem is that it is possible to implement the Fourier
transform over certain groups {\em efficiently} on a quantum computer. 
This in
turn allows to sample the Fourier components efficiently (this
technique is referred to as the ``standard method''). In the case of
Abelian groups $G$ (appearing in factoring and discrete log) the
hidden subgroup can be reconstructed with only a polynomial (in $\log
|G|$) number of queries to the function and a polynomial number of
measurements (samplings in the Fourier basis) and postprocessing
steps.

Addressing the HSP in the non-Abelian case is considered to be one of the most
important challenges at present in quantum computing.  A positive
answer to the question whether quantum computers can 
efficiently solve the Hidden Subgroup Problem over non-Abelian groups
would have several important implications for the solution of problems
in NP, which are neither known to be NP-complete nor in P; and which
are good candidates for a quantum speed-up. Among the most prominent
such problems is Graph Isomorphism, where the group in 
question is the symmetric group.

For many non-Abelian groups it is possible to implement the Fourier transform on a quantum computer efficiently \cite{Ettinger:99a,Roetteler:98a,Pueschel:99a,Grigni:01a,Moore:04b}, and in particular explicit constructions exist for the symmetric group $S_n$ \cite{Beals:97a}. This fact and the prominence of the  problems involved make it very desirable to get a handle on the power of Quantum Fourier Sampling (QFS) to solve the HSP or its decision version for general groups. 

In this paper we focus on the question which hidden subgroups can be distinguished from the identity via QFS with special attention to the symmetric group. Several partial positive results have been obtained previously for groups that are in some ways ``close'' to Abelian, like some semidirect products of Abelian groups \cite{Ettinger:99a,Roetteler:98a,Kuperberg:03a,Moore:04a}, in particular the Dihedral group; Hamiltonian groups \cite{Hallgren:00a}, groups with small commutator groups \cite{Ivanyos:01a} and solvable groups of constant exponent and constant length derived series \cite{Friedl:03a}. Often in these cases the irreducible representations are known and can be analysed in a relatively straightforward way. 
For instance the Dihedral group $D_n$, the first non-Abelian group to
be analysed in this context by Ettinger and Hoyer \cite{Ettinger:99a},
is ``nearly'' Abelian in the sense that all of its irreducible
representations (irreps) have degree at most two. 
Indeed hidden
reflections of $D_n$ can be distinguished from identity with only
polynomial Quantum Fourier Samplings, similar to the Abelian case
(where all irreducible representations are
one-dimensional).\footnote{Note, however, that the computational 
version of HSP seems much harder: even though a polynomial number of
samples suffice to {\em distinguish} hidden reflections {\em
information theoretically}, no efficient reconstruction procedure
beyond sophisticated exhaustive search is known.}

The holy grail of the field is the symmetric group $S_n$, which seems
much harder to analyse, partly because to this day there is still only
partial explicit knowledge about its irreducible representations and
character values \cite{Sagan:book}, because most of its subgroups are
far from normal (have many conjugate subgroups), because most of its
irreducible representations have very large dimension
($2^{\Theta (n \log n)}$) and the number of different irreducible 
representations
is an exponentially small fraction of the size of the group, to
name just some of the difficulties. The structure of distinguishable
versus indistinguishable subgroups of $S_n$ has remained highly
elusive.

In this work we provide a substantial step towards a complete
classification of subgroups of the symmetric group for which the
decision version of the HSP can be solved efficiently via Quantum
Fourier Sampling. We bring into play classical notions and results 
from the theory of finite permutation groups, which have not been
employed before in quantum computing, and seem to be relevant
in these investigations. 
These include notions such as minimal degree, rank,
subdegrees, and primitivity, which played a key role in permutation
group theory since the days of Jordan in the 19th century. Moreover,
recent advances in finite permutation groups, due to sophisticated
work by Babai \cite{Babai:81a, Babai:82a} and others on the one hand,
and due to the Classification of Finite Simple Groups (CFSG) on the
other hand (see Cameron \cite{Cameron:82a}), provide us 
with very powerful machinery. 

Using these notions and machinery we present several new results, 
which incorporate existing 
results in the area, and we provide a toolbox for further investigations. 
We are able to give both upper bounds and, for the first time in this 
context, lower bounds on the total 
variation distance of the relevant distributions, and to derive many
applications. 
\paragraph{Outline of results:}
In a nutshell, our main results show that 
under various conditions on the hidden subgroup $H<S_n$, 
the following statement is true for the hidden subgroup $H<S_n$:
\medskip

\noindent
{\em $\spadesuit$ $H$ can be distinguished from the identity subgroup with either 
the weak standard method or the strong standard method with random basis only 
if it contains an element of constant support (i.e. a permutation in which 
all but a constant number of points are fixed).}
\medskip

Since there is only a polynomial number of such elements in $S_n$, 
the statement implies that the standard method both in its weak and 
strong form with random basis provides no advantage over classical 
exhaustive search.
\begin{itemize} 
\item Statement $\spadesuit$ is true if $H$ is of polynomial size.
\item It is also true for an important class of subgroups known as 
{\it primitive} groups. These subgroups, which can be superpolynomial in size, are considered the building blocks of permutation groups.
\item We exhibit a family of even larger subgroups, of exponential 
order, for which Statement $\spadesuit$  is true.
\end{itemize}
The cases we study seem to suggest that Statement $\spadesuit$ might hold for 
{\it all} subgroups $H$. Surprisingly we show that this is indeed true 
under a plausible group theoretic conjecture, for which we provide further evidence.  Assuming the conjecture, 
it follows that Quantum Fourier Sampling (with  random  basis) 
provides no advantage whatsoever over classical exhausitve search.

\paragraph{Main results:} 
We focus on the {\em weak} form of the standard method (see Section \ref{s:2}), since the
strong form with random choices of basis does not provide any
non-negligible additional information for the symmetric group and the
subgroups we consider \cite{Grigni:01a}\footnote{
The strong standard method sometimes
provides substantially more information than its weak counterpart, and
is indeed necessary to efficiently solve HSP in the case of groups
like the Dihedral group \cite{Ettinger:99a,Kuperberg:03a} and other semidirect
product groups \cite{Moore:04a}. An irrep, and hence the non-Abelian
QFT, is given only up to the choice of basis. Grigni et al. show that
for a {\em random} basis the additional information provided by the
strong method is exponentially small, provided the group is
sufficiently non-Abelian and the hidden subgroup sufficiently small,
as is the case for all groups we analyse here. It remains to be seen
whether judicious choices of basis for each irrep can give more
information in the case where  random choices don't help; but this is
believed to be unlikely.}. 
In particular we are not aware of any cases where 
there is a good basis for the strong method, but a {\em random} choice of basis does not also solve the HSP.

To state our main results,
let $G$ be a finite group, $H \le G$ a subgroup, and let $D_H$ denote 
the total variation distance between the distributions on the irreducible
representations of $G$ induced by $H$ and by the identity respectively, 
when sampled with the weak standard method. We say that $H$ is {\it
distinguishable} (using the weak standard method) if $D_H \ge
(\log{|G|})^{-c}$ for some constant $c$, and is {\it indistinguishable}
otherwise.
 
Our starting point is a general result providing both upper and 
lower bounds on the total variation distance $D_H$ in terms of the 
same group theoretic data.

\begin{thm}\label{Th:main}
Let $C_1, \ldots, C_k$ denote the non-identity conjugacy classes of $G$. 
Then 
\begin{equation}
\sum_{i=1}^k |C_i \cap H|^2 |H|^{-1}|C_i|^{-1}< D_H \le 
\sum_{i=1}^k |C_i \cap H||C_i|^{-\frac{1}{2}}.
\end{equation}
\end{thm}
Applying the upper bound with $|H|=2$ gives the result obtained previously by  Hallgren et al. and Grigni et al. \cite{Hallgren:00a,Grigni:01a}.
No lower bounds seem to exist in the literature.
This theorem has a wide range of applications. 
For example, it enables us to
characterise distinguishable subgroups $H \le G$ of polylogarithmic order
(see Theorem \ref{Th:polygeneral} below).

Specialising to $G = S_n$ we show that the minimal degree of $H$
is a crucial notion in the study of the distinguishability of $H$.
The {\it minimal degree} $m(H)$ of a permutation group $H$ 
is defined to be the minimal number of points moved by a non-identity
element of $H$. In other words, for $g \in S_n$ let $fix(g)$ be
the number of fixed points of $g$, and let $supp(g) = n-fix(g)$ be the support of $g$.
Then
\[
m(H) = \min \{ supp(h): 1 \ne h \in H \}.
\]

This notion goes back to the 19th century, and plays an important
role in the theory of finite permutation groups since the days
of Jordan \cite{Jordan:73a,Jordan:75a}. 
It is intriguing that it plays some
role in the HSP as well, 
giving a complete characterisation of distinguishable polynomial 
size subgroups:

\begin{thm} \label{Th:poly}
Let $H \le S_n$ with $|H| \le n^c$ for some constant $c$. 
Then $H$ is distinguishable if and only if its minimal degree $m(H)$ 
is constant. 
\end{thm}

For instance we cannot distinguish a group generated by a cycle of non-constant length or an involution with non-constant number of transpositions (implying the result in \cite{Hallgren:00a,Grigni:01a}). Note that the strength of this theorem comes from the ``if-and-only-if'': The  distinguishable subgroups must contain an element of constant support. Since there are only polynomially many such elements in $S_n$ we can just exhaustively query them. 

This also has implications for the Graph Isomorphism (GI) problem. Recall that to solve GI for two graphs $G_1,G_2$, it suffices to distinguish a hidden subgroup of the automorphism group $Aut(G_1 \cup G_2)$ of the form $H_1 \times H_2$ (not $G_1 \simeq G_2$), where $H_i=Aut(G_i)$, from a subgroup of the form $H \cup \sigma H$ ($G_1 \simeq G_2$), where $H=H_1 \times H_2$ and $\sigma$ maps $G_1$ to $G_2$ (see e.g. \cite{Josza:00a}). If the automorphism group of each graph is of polynomial size our results imply that we cannot distinguish each of the two possible cases from identity, and hence (using the triangle inequality) we cannot distinguish them from each other unless Aut($G_i$) contains an element of constant support. Thus QFS provides no advantage here. 

Our next result concerns {\em primitive} subgroups. A permutation group is called {\it primitive} if it is
transitive (has only one orbit) and does not preserve a non-trivial partition (block system)
of the permutation domain. Primitive permutation groups are considered
the building blocks of finite permutation groups in general, and 
were extensively studied over the past 130 years. We note that if
$H \le S_n$ is primitive and $H \ne A_n, S_n$ then Babai showed that
$m(H) \ge (\sqrt{n}-1)/2$ and $|H| \le n^{4\sqrt{n}\log{n}}$. Using
the Classification of Finite Simple Groups the latter bound can be 
somewhat improved  to $|H| \le 2n^{\sqrt{n}}$, which is essentially
best possible \cite{Cameron:82a}; in particular the order of $H$ can 
be much more than polynomial, 
and so Theorem \ref{Th:poly} above does not apply.

However, we obtain the following somewhat surprising general result:

\begin{thm}\label{Th:primitive}
Let $H \ne A_n, S_n$ be a primitive subgroup.
Then $H$ is indistinguishable.
\end{thm}

As the hidden subgroups get large we would suppose that it becomes easier 
to distinguish them from the identity. 
However, we show below that $H$ can get extremely large and yet cannot 
be distinguished with the weak standard method:

\begin{thm}\label{Th:large}
Let $\epsilon(n)$ be a sequence of real numbers which tend to
zero as ${n \rightarrow \infty}$. 
Then for all sufficiently large $n$ there is an indistinguishable 
subgroup $H < S_n$ of size $|H| \ge |S_n|^{\epsilon(n)}$. 
\end{thm}

In particular, there are indistinguishable subgroups $H$ of 
exponential order.

To prove Theorems \ref{Th:primitive}, \ref{Th:large} we give a
somewhat technical group theoretic criterion for indistinguishibility
of subgroups of non-constant minimal degree (Proposition \ref{Prop}).
We conjecture that this criterion applies universally.  This implies
that every distinguishable subgroup has a non-identity element of
constant support.

It is interesting that permutation-theoretic data is relevant 
to the distinguishability problem even when the group $G$ in
question is not $S_n$, but an arbitrary finite group. 
Indeed, given $H \le G$ there is a standard way to view $G$ 
(or $G/N$ where $N$ is the normal core of $H$) as 
a transitive permutation group on the set $X = G/H$ of (right) 
cosets of $H$ in $G$ (where $g \in G$ acts by right multiplication).
Recall that a {\it suborbit} of $G$ in this action is an orbit of 
$H$ on $X$, and the {\it rank} $r_X(G)$ of $G$ is defined to be 
the number of suborbits of $G$. The {\it subdegrees} of $G$ are
the sizes of its suborbits. 
Thus the average subdegree of $G$ is $|G:H|/r_X(G)$.  
Setting $H^g = gHg^{-1}$ it is easy to see that 
the subdegrees of $G$ have the form $|H:H \cap H^g|$ 
for $g \in G$.
Using this data, Theorem \ref{Th:main}, and classical permutation theoretic
tools, we obtain the following positive result.

\begin{thm} \label{Th:orbit}
Suppose $|H|$ is not polylogarithmic, but the average subdegree 
of $G$ on $G/H$ is polylogarithimic.
Then $H$ is distinguishable.
In particular this holds when $|H:H \cap H^g| \le (\log{|G|})^c$
for all $g \in G$.
\end{thm}

This theorem extends the result of \cite{Hallgren:00a} showing that if $H$
is normal in $G$ then $H$ is distinguishable (since in this case 
$|H:H \cap H^g|=1$ for all $g$). It also implies the easy observation 
that subgroups of size at least $|G|/{(\log{|G|})^c}$ are always 
distinguishable.

Our methods also allow us to examine the more general case of
distinguishing between two subgroups $H$ and $K$ of $G$, see 
Section 3 for some details.

\paragraph{Further Related Work:}
The HSP plays a central role in most known quantum algorithms and the efficient algorithm for the Abelian case using Fourier Sampling is folklore. The non-Abelian HSP has received a lot of attention in recent years, due to its connection to several candidate problems in NP like Graph Isomorphism (for the symmetric group) and lattice problems \cite{Regev:02a} (for the Dihedral group); we mention only the work relevant to ours.

Despite a lot of progress for various non-Abelian groups \cite{Ettinger:99a,Roetteler:98a,Moore:04a,Ivanyos:01a,Friedl:03a} the results on the symmetric group are very sparse. Grigni et al. \cite{Grigni:01a} show that sampling the row index in the strong standard method provides no additional information. 
They also show that the additional information provided by the strong method in a {\em random} basis scales with $\sqrt[3]{|H|^2 k(G)/|G|}$ where $k(G)$ is the number of conjugacy classes of the group $G$ and $|H|$ the size of the hidden subgroup. Both Hallgren et al. and Grigni et al. \cite{Hallgren:00a,Grigni:01a} show that hidden subgroups of $S_n$ of size $|H|=2$, generated by involutions with large support, cannot be distinguished from identity; exactly the task that needs to be solved for Graph Automorphism.

Hallgren et al. \cite{Hallgren:00a} also point out that the weak standard method cannot distinguish between conjugate subgroups. In \cite{Hallgren:00a,Grigni:01a} it is shown that the weak standard method allows us to efficiently determine the normal core of a hidden subgroup $H$ and hence in particular normal subgroups.

\section{Preliminaries and notation}\label{s:2}

Fix a finite group $G$ and a subgroup $H \le G$. We denote states of the vector space $\mathbb{C}[G]$, spanned by the group elements, with a $|\cdot\ra$, as is standard in quantum computation\footnote{For the necessary background in quantum computation see e.g. \cite{Nielsen:book}.}.

The Quantum Fourier Transform (QFT) over a group $G$ is the following unitary transformation on $\mathbb{C}[G]$:
$$|g\ra \rightarrow \frac{1}{\sqrt{|G|}} \sum_{\rho,i,j} \sqrt{d_\rho} \rho(g)_{ij} |\rho,i,j\ra$$ where $\rho$ labels an irreducible representation of $G$, $d_\rho$ is its dimension and $1 \leq i,j \leq d_\rho$. The $|\rho,i,j\ra$ span another basis of $\mathbb{C}[G]$, the so called Fourier basis. 

The {\it standard method} of Quantum Fourier Sampling is the following: The state is initialised in a uniform superposition over all group elements; a second register is initialised to $|0\ra$. Then the function $f$ is applied reversibly over both registers (i.e. $f:|g\ra |0\ra \rightarrow |g\ra |f(g)\ra$). Finally the second register is measured, which puts the first register  into the superposition of a (left)-coset of $H$, i.e.  in the state $|gH\ra:=\frac{1}{\sqrt{|H|}} \sum_{h \in H} |gh\ra$ for some random $g \in G$. Finally the QFT over $G$ is performed, yielding the state
$$\frac{1}{\sqrt{|G||H|}} \sum_{\rho,i,j} \sqrt{d_\rho} \sum_{h \in H} \rho_{ij}(gh)|\rho,i,j\ra.$$
A basis measurement now gives $(\rho,i,j)$ with probability $P_{gH}(\rho,i,j)=\frac{d_\rho}{|G||H|}|\sum_{h \in H}\rho_{ij}(gh)|^2.$

Since we do not know $g$ and $g$ is distributed uniformly, we sample $(\rho,i,j)$ with probability $P_{H}=\frac{1}{|G|}\sum_{g} P_{gH}$.
The {\it strong} standard method samples both $\rho$ and its entries $i,j$.  In the {\it weak} standard method  only the character $\chi_\rho$ is measured  (but not the entries $i,j$, which are averaged over)\footnote{It is easy to see \cite{Hallgren:00a,Grigni:01a} that for the weak standard method the probability to sample $\rho$ is independent of the coset of $H$ we happen to land in.}. The probability to measure $\rho$ in the weak case is $$P_H(\rho)=\frac{d_\rho}{|G|}\sum_{h \in H} \chi_\rho(h).$$
Let $Irr(G)$ be the set of irreducible 
characters of $G$. Then $P_H$ is a distribution on $Irr(G)$.

To solve HSP we need to infer $H$ from the resulting distribution. Distinguishing the trivial subgroup $\{e\}$ from a larger subgroup $H$ 
efficiently using the standard method is possible if and only if the $L_1$ distance $D_H$ between $P_{\{e\}}$ and $P_H$ is larger than some inverse polynomial in $\log |G|$. The $L_1$ distance (also known as the total variation
distance) is given as
\begin{equation}\label{DH}
D_H=\frac{1}{|G|} \sum_\rho d_\rho |\sum_{h \in H, h \neq e} \chi_\rho(h)|.
\end{equation}
We say that $H$ is {\it distinguishable} (using the weak standard method)
if $D_H \ge (\log{|G|})^{-c}$ for some constant $c$, and {\it indistinguishable} otherwise. 
If $K \le G$ is another subgroup
we let $D(H,K) = |P_H-P_K|_1$ be the $L_1$ distance between the 
distributions $P_H$ and $P_K$.

We also need some group theoretic notation. For $x \in G$ we let
$x^G$ denote the conjugacy class of $x$ in $G$.
Let $C_1, \ldots , C_k$ denote the non-identity conjugacy classes of $G$.
For an irreducible character $\chi_{\rho} \in Irr(G)$ we let
$\chi_{\rho}(C_i)$ denote the common value of $\chi_{\rho}(x)$
for elements $x \in C_i$.

\section{Arbitrary groups}\label{s:3}

In this section we discuss results for arbitrary groups $G$,
providing some proofs when space allows.
\medskip

\noindent
{\it Proof of Theorem \ref{Th:main}.}
For each irreducible representation $\rho$ of $G$ we have
\[
|\sum_{h \in H, h\neq e} \chi_\rho(h)| \le  
\sum_{h \in H, h\neq e} |\chi_\rho(h)| \le  
\sum_{h \in H, h\neq e} d_\rho < |H|d_{\rho}.
\]
Hence 
$d_{\rho} > |H|^{-1}|\sum_{h \in H, h\neq e} \chi_\rho(h)|$.
Substituting this in (\ref{DH}) we obtain
\[
D_H > \frac{1}{|G||H|}
\sum_{\rho} |\sum_{h \in H, h\neq e} \chi_\rho(h)|^2.
\]
Note that $\chi_\rho(h) = \chi_{\rho}(C_i)$ 
if $h \in H \cap C_i$. This yields
$\sum_{h \in H, h\neq e} \chi_\rho(h) =  
\sum_{i=1}^k |H \cap C_i| \chi_\rho(C_i)$,
and so
\[
D_H > \frac{1}{|G||H|}\sum_{\rho}
|\sum_{i=1}^k |H \cap C_i| \chi_\rho(C_i)|^2.
\]
Now,
\[
|\sum_{i=1}^k |H \cap C_i| \chi_\rho(C_i)|^2 =
\sum_{i=1}^k |H \cap C_i|^2 |\chi_\rho(C_i)|^2 +
\sum_{i \ne j} |H \cap C_i||H \cap C_j| 
\chi_\rho(C_i) {\bar{\chi}_{\rho}}(C_j).
\]
Using the generalised orthogonality relations we observe that
\[
\sum_{\rho} \sum_{i=1}^k |H \cap C_i|^2 |\chi_\rho(C_i)|^2 
= \sum_{i=1}^k |H \cap C_i|^2 |G|/|C_i|,
\]
and 
\[
\sum_{\rho} \sum_{i \ne j} |H \cap C_i||H \cap C_j| 
\chi_\rho(C_i) {\bar{\chi}_{\rho}}(C_j) = 0.
\]
It follows that
\[
D_H > \frac{1}{|G||H|} \sum_{i=1}^k |H \cap C_i|^2 |G|/|C_i| =
\sum_{i=1}^k  |H \cap C_i|^2 |H|^{-1}|C_i|^{-1}.
\]
This completes the proof of the lower bound.
To prove the upper bound, write
\begin{equation}\label{Eq:upperDH}
D_H |G|=\sum_{\rho} d_\rho |\sum_{h \in H, h\neq e} \chi_\rho(h)| \le
\sum_{\rho} d_\rho \sum_{h \in H, h\neq e} |\chi_\rho(h)| = 
\sum_{h \in H, h\neq e} \sum_{\rho} d_{\rho} |\chi_\rho(h)|.
\end{equation}

Fix $h \in H$ and choose $i$ such that $h \in C_i$.
Using the Cauchy-Schwarz inequality we obtain
\[
\sum_{\rho} d_{\rho} |\chi_\rho(h)| \le 
(\sum_{\rho} d_{\rho}^2)^{1/2} (\sum_{\rho} |\chi_{\rho}(h)|^2)^{1/2},
\]
giving (using the orthogonality relations)
$\sum_{\rho} d_{\rho} |\chi_\rho(h)| \le |G|^{1/2}(|G|/|C_i|)^{1/2} = 
|G||C_i|^{-1/2}$.
Summing over non-identity elements $h \in H$, and observing that
the upper bound above occurs $|H \cap C_i|$ times, we obtain
$\sum_{h \in H, h \ne e} \sum_{\rho} d_{\rho} |\chi_\rho(h)| \le 
\sum_{i=1}^k  |H \cap C_i||G||C_i|^{-1/2}$.
Combining this with 
(\ref{Eq:upperDH}) 
we obtain
$D_H \le \sum_{i=1}^k |H \cap C_i| |C_i|^{-1/2}$,
as required.
\qed
\medskip

The following is an immediate consequence of Theorem \ref{Th:main}.

\begin{cor} \label{corr:1}
Let $C_{min}$ denote a non-identity conjugacy class of minimal 
size intersecting $H$ non-trivially . Then we have
\[
|H|^{-1}|C_{min}|^{-1} < D_H \le (|H|-1)|C_{min}|^{-1/2}.
\]
\end{cor}

We can now characterise distinguishable subgroups of polylogarithmic 
order in an arbitrary group $G$. Assuming $|H|$ is polylogarithmic 
Corollary \ref{corr:1} shows that $D_H^{-1}$ is polylogarithmic
if and only if $|C_{min}|$ is. In other words we have proved the
following.

\begin{thm}\label{Th:polygeneral}
Suppose $|H| \le (\log{|G|})^c$ for some constant $c$. 
Then $H$ is distinguishable if and only if $H$ has a non-identity element 
$h$ such that $|h^G| \le (\log{|G|})^{c'}$ for some constant $c'$.
\end{thm}

There is an interesting reformulation of the lower bound in  
Theorem \ref{Th:main} in terms of fixed points. Regard $G$ as a 
permutation group on $X = G/H$. Denote by $fix_X(g)$ the number of fixed
points of $g \in G$ in this action. Let $r = r_X(G)$ denote the
rank of $G$ in this action, namely, the number of orbits of
the point stabilizer $H$ on $X$.

\begin{cor} With the above notation we have

(i)$D_H > {\frac{1}{|G|}} \sum_{h \in H, h \ne e} fix_X(h)$;

(ii) $D_H > r_X(G)/|X| - 1/|H|$.

\end{cor}

\paragraph{Proof:}
It is well known that, if
$g \in C_i$, then 
$fix_X(g)/|X| = |H \cap C_i|/|C_i|$.

Therefore
\[
\sum_{i=1}^k  |H \cap C_i|^2 |H|^{-1}|C_i|^{-1} =
|H|^{-1} \sum_{h \in H, h \ne e} fix_X(h)/|X| =
|G|^{-1} \sum_{h \in H, h \ne e} fix_X(h).
\]
Combining this with Theorem \ref{Th:main} we deduce part (i).

To prove part (ii) we use the well known lemma of Frobenius
that the number of orbits of $H$ on $X$ equals the average number
of fixed points of $h \in H$ on $X$ (see for instance Theorem 1.7A
of \cite{Dixon:96a}). This shows that
\[
|H|^{-1} \sum_{h \in H, h \ne e} fix_X(h) = r_X(G) - |X|/|H|.
\]
Dividing both sides by $|X|$ we deduce part (ii) from
part (i).
\qed
\medskip

\noindent
{\it Proof of Theorem \ref{Th:orbit}.}
This follows easily from part (ii) of the above Corollary.
\qed
\medskip

We close this section by considering the more general problem of 
distinguishing between two arbitrary subgroups $H$ and $K$ of $G$. 
Obviously, if the total variation distance $D(H,K)$ between the
respective distributions is zero then the weak standard method cannot 
distinguish between $H$ and $K$, even if superpolynomial complexity is
allowed. This gives rise to the fundamental problem of characterising
subgroups $H, K$ of distance zero. It has already been observed
that conjugate subgroups have distance zero \cite{Hallgren:00a}, 
does the converse hold?

To solve these problems, recall that the permutation representation 
of $G$ on $G/H$ gives rise to a linear representation of $G$ in dimension 
$G/H$ (which can be realised using the corresponding permutation matrices). 
We can now state

\begin{thm} The following are equivalent for subgroups $H,K \le G$.

(i) $D(H,K)=0$.

(ii) For each conjugacy class $C$ of $G$ we have $|H \cap C| = |K \cap C|$.

(iii) The permutation representations of $G$ on $G/H$ and
$G/K$ give rise to equivalent linear representations.

Moreover, there exist finite groups $G$ and non-conjugate subgroups
$H,K \le G$ such that $D(H,K)=0$.
\end{thm}

Our proof of the equivalence of (i)-(iii) is elementary, based
on the fact that the characters in $Irr(G)$ form a base for the
class functions on $G$. The proof of the last assertion is deeper
and will be omitted in this version.

\section{Symmetric groups} \label{s:4}

Let us now focus on the case $G = S_n$.
\medskip

\noindent
{\it Proof of Theorem \ref{Th:poly}.}
Let $g \in S_n$ with $supp(g)=k$. Then it is straightforward to verify 
that $\binom {n}{k} \le |g^{S_n}| \le n^k$.
As a consequence we see that a conjugacy class $C$ in $S_n$ 
has polynomial order if and only if it consists of elements
of constant support. This observation, when combined with
Theorem \ref{Th:polygeneral}, completes the proof of Theorem \ref{Th:poly}.
\qed
\medskip

The proofs of Theorems \ref{Th:primitive} and \ref{Th:large} are longer 
and less elementary, and so we will only sketch them here. In the heart
of both proofs lies the following somewhat technical result.

\begin{prop} \label{Prop}
Let $H \le  S_n$ be a subgroup with non-constant minimal degree. Suppose
that, for each $k \le n$, $H$ has at most $n^{k/7}$ elements of support 
$k$. Then $H$ is indistinguishable.
\end{prop}

\paragraph{Proof:}
Apply the upper bound of Theorem \ref{Th:main}, written in the form
\[
D_H \le \sum_{1 \ne h \in H} |h^G|^{-1/2}.
\]
To evaluate this sum we use a result from \cite{Liebeck:01a},
showing that, for $G=S_n$ and for $h \in G$ of support $k$
we have $|h^G| > n^{ak}$ for any real number $a < 1/3$ and $n$ 
large enough (given $a$).
Setting 
\[
H_k = \{ h \in H: supp(h)=k \}, 
\]
we obtain
\[
D_H < \sum_{k \ge m(H)} |H_k|n^{-bk},
\]
for any real number $b < 1/6$ and sufficiently large $n$.
Fix $b$ with $1/7 < b < 1/6$, and set $c = b-1/7$, $m=m(H)$.
Then
\[
D_H < \sum_{k \ge m} n^{k/7}n^{-bk} = \sum_{k \ge m} n^{-ck} \le 2n^{-cm}.
\]
Since $m=m(H)$ is non-constant, we see that $D_H$ is smaller than
than any fixed negative power of $n$, and so $H$ is indistinguishable.
\qed
\medskip

Now, for Theorem \ref{Th:primitive} we use Babai's lower bound on the 
minimal degree of primitive subgroups $H \ne A_n, S_n$ \cite{Babai:81a}, 
showing that 
\begin{equation}\label{Eq:B}
m(H) \ge (\sqrt{n}-1)/2.
\end{equation}
Furthermore, we apply a theorem of Cameron \cite{Cameron:82a}
(which in turns relies on the Classification of Finite Simple Groups) 
describing all primitive groups of `large' order. In particular it
follows from that description that, for all large $n$, and for a primitive
subgroup $H \ne A_n,S_n$, either

(i) $|H| \le n^{c n^{1/3}}$, or

(ii) $n = \binom{l}{2}$ for some $l$, and $H \le S_l$ acting
on $2$-subsets of $\{1, \ldots , l \}$, or

(iii) $n=l^2$ for some $l$, and $H \le S_l \wr S_2$ acting on 
$\{ 1, \ldots , l \}^2$ in the so called product action.

We claim that for all large $n$ and for all $k$ we have $|H_k| \le n^{k/7}$. 

To show this it suffices to consider $k \ge (\sqrt{n}-1)/2$, otherwise 
$|H_k|=0$ by (\ref{Eq:B}).
Now, if $H$ satisfies condition (i) above then the claim follows trivially
using $|H_k| \le |H|$. So it remains to consider groups $H$ in cases 
(ii) and (iii). Here a detailed computation based on the known actions of 
$H$, which we omit from this version, completes the proof of
the claim.

At this point Proposition \ref{Prop} can be applied, and we conclude
that $H$ is indistinguishable.
In fact our argument shows that, for some $\epsilon > 0$, all primitive 
subgroups $H \ne A_n,S_n$ satisfy $D_H < n^{-\epsilon \sqrt{n}}$.

Finally, to prove Theorem \ref{Th:large} we construct $H$ as the full
symmetric group on $\lfloor n/r \rfloor$ blocks of size $r$, where $r = r(n)$ tends
very slowly to infinity. Then $m(H)=2r$, which is non-constant.  A
detailed computation  shows that the second assumption of
Proposition \ref{Prop} also holds, which yields the desired
conclusion.  The details are left to the reader.
\medskip

We end this paper with 
\medskip

\noindent
{\bf Conjecture $\spadesuit$:} {\it Suppose $H \le S_n$ is distinguishable.
Then its minimal degree $m(H)$ is constant.}
\medskip

\noindent
{\bf Conjecture $\clubsuit$:} {\it Every subgroup $H \le S_n$ with non-constant minimal degree has at most $n^{k/7}$ elements of support $k$.}

Proposition \ref{Prop} shows that Conjecture $\clubsuit$ implies Conjecture $\spadesuit$. We regard Conjecture $\clubsuit$ as a plausible group theoretic conjecture, for which we have mounting evidence.

First note that by the above discussion  Conjecture $\clubsuit$ holds for primitive groups. Secondly we can show that the set of subgroups satisfying the conjecture is closed under direct products. Thirdly we can prove the conjecture for wreath products $K \wr L$, if $K$ satisfies the conjecture. 
Recall that all 
transitive imprimitive groups are subgroups of wreath products $W=S_k\wr S_l$ and the {\em maximal} ones are the full wreath product $W$. 

Hopefully the methods introduced in this paper and the group theoretic reductions will lead to a full classification of distinguishable subgroups of $S_n$ and of other groups.

\section*{Acknowledgement} JK thanks Oded Regev for helpful comments on an earlier version of this paper. JK's effort was supported by ACI S\'ecurit\'e
Informatique, 2003-n24, projet "R\'eseaux Quantiques", ACI-CR 2002-40
and EU 5th framework program RESQ IST-2001-37559 as well as by the
Defense Advanced Research Projects Agency (DARPA) and the Air Force
Laboratory, Air Force Material Command, USAF, under Contract
No. F30602-01-2-0524 and by the National Science Foundation through
Grant No. 0121555.



\newcommand{\etalchar}[1]{$^{#1}$}

\end{document}